\documentclass[AMA,STIX1COL,doublespace]{WileyNJD-v2}

\articletype{}%

\received{7 May 2018}
\revised{31 August 2018}
\accepted{}

\usepackage{placeins}

\usepackage{fancyvrb}
\usepackage{url}
\usepackage{algpseudocode}
\usepackage{algorithm}

\begin{document}

\title{Tuning the Performance of a Computational Persistent Homology Package}

\author[1]{Alan Hylton}
\author[2]{Gregory Henselman-Petrusek}
\author[3]{Janche Sang}
\author[4]{Robert Short}

\address[1]{\orgdiv{Space Communications and Navigation}, \orgname{NASA Glenn Research Center}, \orgaddress{\state{Cleveland, OH}, \country{USA}}}

\address[2]{\orgdiv{Dept. of Electrical and Systems Engineering}, \orgname{University of Pennsylvania}, \orgaddress{\state{Philadelphia, PA}, \country{USA}}}

\address[3]{\orgdiv{Dept. of Elect. Eng. and Computer Science}, \orgname{Cleveland State University}, \orgaddress{\state{Cleveland, OH}, \country{USA}}}

\address[4]{\orgdiv{Dept. of Mathematics}, \orgname{Lehigh University}, \orgaddress{\state{Bethlehem, PA}, \country{USA}}}

\abstract[Summary]{In recent years, persistent homology has become an attractive
method for data analysis. It captures topological features, such as 
connected components, holes, and voids from point cloud data and summarizes
 the way in which these features appear 
and disappear in a filtration sequence. In this project, we focus on improving
the performance of Eirene, a computational package for persistent homology. 
Eirene is a 5000-line open-source software library implemented in the 
dynamic programming language Julia. We use the Julia profiling tools to
identify  performance bottlenecks and develop novel methods to manage them, 
including the parallelization of some time-consuming
functions on  multicore/manycore hardware.  Empirical results show
that performance can be greatly improved. }


\maketitle

\section{Introduction}

Persistent Homology is a mathematical model for shape description that is specially adapted, in an algorithmic context, to    
challenges in data analysis  that arise from noise, bias, dimensionality, and curvature.  It has given rise, over the past three decades, to a highly productive branch of data science known as Topological Data Analysis [TDA], with prominent works including \cite{FrosiniMeasuring91, RobinsTowards99, PHomology, ZCComputing05, Chazal:2009:PPM:1542362.1542407, Cohen-Steiner2007, lesnick2011}.   Excellent surveys may be found in a number of sources, such as \cite{CarlssonTopology09, GhristBarcodes08, EMPersistent12, Giusti2016}.  The key ingredients to this model are \emph{homology}, a sequence of topological shape descriptors that include graph theoretic constructs (connectivity and corank) and more general structures commonly characterized as holes or voids, and \emph{functoriality}, which relates  the homologies of different shapes.  A combination of these ingredients results in persistent homology:  a unified mathematical framework for analyzing the evolution of homological features that change over time.  The advantages of this model derive from the advantages of homological shape descriptors, namely robustness to deformation and blindness to ambient dimension, and from the rich body of literature surrounding functorial shape comparison.
Persistent homology finds numerous applications 
in research areas that involve  large data sets, including 
biology \cite{PHevolution}, image analysis \cite{PHimage},
sensor networks \cite{PHwsn}, Cosmology \cite{PHcosmology}, etc.
A detailed roadmap for the computation of persistent homology can 
be found in \cite{OtterRoadMap}. 

Eirene is an open-source platform for computational
persistent homology \cite{Eirene}. It is implemented in Julia \cite{Julia},
a high-performance dynamic scripting language for numerical and 
scientific computation. Eirene exploits the novel relationship 
between  Schur complements (in linear algebra), discrete
Morse Theory (in computational homology), and minimal bases (in combinatorial
optimization) described in \cite{HenselmanThesis} to augment the standard
algorithm for homological persistence computation \cite{ZCComputing05}.  This optimization has been
shown, in certain cases, to improve the performance of the standard
algorithm by several orders of magnitude in both time and memory.  The library additionally
includes built-in utilities for graphical visualization of homology 
class representatives.

However, we noticed that Eirene ran slower with the recent
release Julia v0.6. Therefore, the objective of this project was 
to optimize the performance of Eirene. We used a profiling technique to
identify  potential performance bottlenecks. Note that software 
profiling, which can display the call graph and the amount of time 
spent in each function, has been used to tune program performance
for several decades\cite{GrahamKM83}\cite{McKusick}. After locating
each bottleneck, we found the cause of it 
and developed a solution. For certain time-consuming
functions, we re-implemented the code and ported it on
 multicore and manycore architectures,  using pthreads \cite{NBF}
and CUDA threads \cite{KirkHwu}, respectively. 
Experimental results show that the performance can be improved
significantly.

This article is an enhanced version of the paper presented at the 
IEEE IPCCC conference~\cite{EireneIPCCC}.
The rest of the paper is organized as follows. In Section 2, we
briefly review the necessary background on persistent homology. 
An example of using Eirene is demonstrated in Section 3.
In Section 4, we identify
 bottlenecks and propose performance-improving methods.
We evaluate our methods by conducting benchmark experiments in Section 5.
A short conclusion is given in Section 6.    

\section{Background}

\begin{figure}
\centerline{}
\centerline{\includegraphics[width=5.75in]{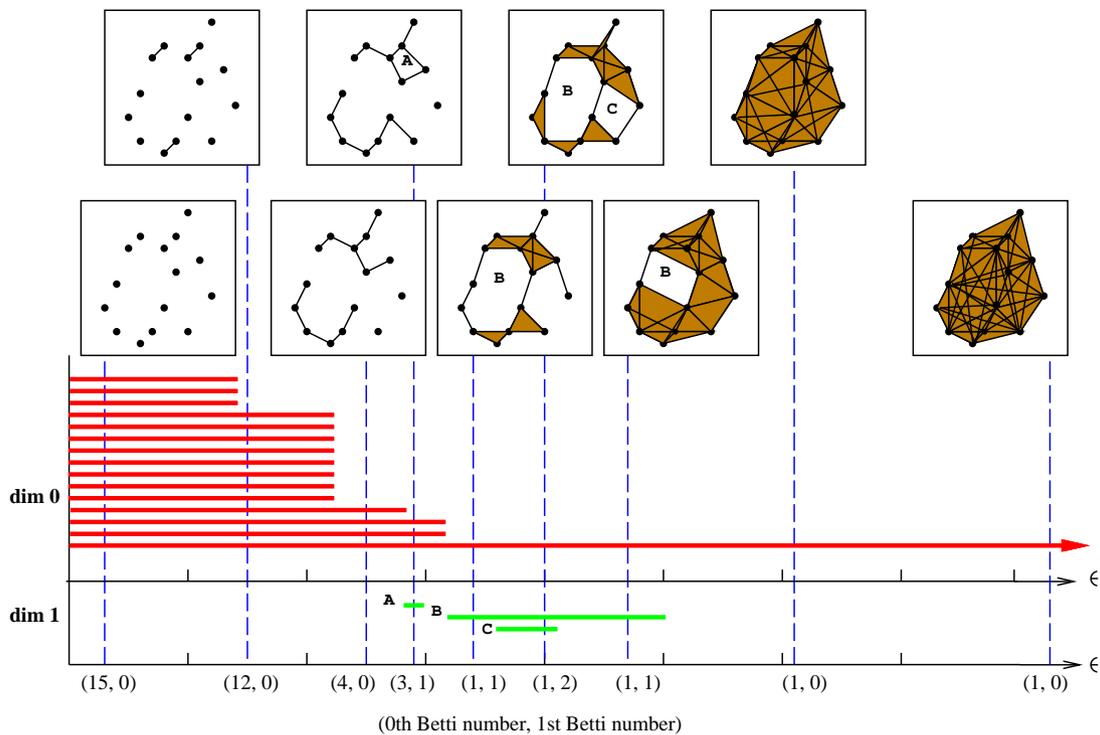}}
\caption{An example of zero- and one-dimensional barcodes for a 
sequence of Vietoris-Rips complexes}
\label{Barcodes}
\centerline{}
\centerline{}
\end{figure}

 Homology is a tool in algebraic topology
used to analyze the connectivity of simplicial complexes, such as  points,
edges, solid triangles, solid tetrahedra, and other higher dimensional
shapes. By using homology, we can measure several features of   
the data in metric spaces -- including the numbers of  connected 
components (0-cycles), 
holes (1-cycles), voids (2-cycles), etc. We limit the scope of this project  to the 
homology of filtered Vietoris-Rips complexes only. 

\subsection{Homological Persistence} \label{subsec:homologicalpersistence}
To construct a filtered Vietoris-Rips complex, 
a distance threshold $\epsilon$ is chosen first. 
Then, any two points that are less than the distance 
$\epsilon$ from each other are connected by an edge. A solid triangle
is created if all its three edges have been generated. A solid
tetrahedra is constructed if all its face triangles have been created.      
Similar constructions are used to build higher-dimensional simplices.

Note that  different distance thresholds will generate different
 Vietoris-Rips complexes, and may therefore  
result in very different homologies, i.e different
numbers and types of cycles. Because cycles are the topologically 
significant features to be captured from the data set,
we do not want to miss any of them. For solving this problem,
a method, which is called persistent homology \cite{PHomology}, is to
generate Vietoris-Rips complexes from a set of points at every distance
threshold and then derive when the cycles appear and disappear
in the complexes as the distance threshold increases. Therefore, the distance
threshold is often referred to as time. 
 
The topological data produced by using persistent homology
can be visualized through a barcode \cite{barcodes}. A barcode is a collection
of intervals, where each interval represents a homology class that
exists in at least one Vietoris-Rips complex generated from a 
point cloud.  The left and the right endpoint of an 
interval represents the birth and the death times of a class, 
respectively. The horizontal axis corresponds to the 
distance threshold and the vertical axis represents an 
arbitrary ordering of captured classes. 

Figure~\ref{Barcodes} illustrates
an example of zero- and one-dimensional barcodes for a 
sequence of Vietoris-Rips complexes. The red bars represent 
the lifetime of the connected components. Note that in the beginning 
the number of components is the same as the number of points.
When the distance threshold $\epsilon$ increases, the number of 
components is decreased because more and more components
are connected together. The green bars represent the lifespans of 
the cycles in dimension 1. A cycle disappears when it is completely 
filled in by solid triangles. It can be observed that
one cycle, denoted as B in the diagram, is significantly longer than
the others, while the cycle A is relatively short-lived.  Short-lived 
cycles are often regarded as ``noise,'' and long-lived cycles as 
``true signal,'' though these interpretations are subjective in 
nature and their fitness depends heavily on context. 
A barcode also encodes all the information regarding 
the Betti numbers on different scales. Betti numbers  
count topological features, like connected 
components (0th Betti number), holes (1st Betti number), 
voids (2nd Betti number), etc., for an individual simplicial complex. 
The Betti numbers of the Vietori-Rips complex with distance 
threshold $\epsilon$ can be found 
by counting the number of intersections between the barcode and
the vertical line the passes through $\epsilon$ on the $x$ axis.  
The 0th and 1st Betti numbers for several different choices of $\epsilon$
appear along the $x$-axis in Figure~\ref{Barcodes}.

For the computation of persistent homology,
there are several open-source software packages available, 
such as javaPlex \cite{Javaplex}, Dionysus \cite{Dionysus}, Perseus \cite{Perseus}, Ripser \cite{ripserwebsite}, and several others 
\cite{
dipha,
Bauer2014,
DBLP:journals/corr/BubenikD15,
2016arXiv160907517D,
dlotckopltoolbox,
fasyrtda,
2014arXiv1411.1830F,
lesnickrivet,
maria2014gudhi,
deysimpers,
deygicomplex,
perryplex,
TVAjavaPlex12}.   Understanding the performance of persistence solvers is a famously complicated problem.  Many elements add to this challenge, but a key contributor is the tremendous variability to be found in  size, complexity, and formatting of the data passed as input.  A good discussion of the benchmarking problem, and of cross-input variation, may be found in  \cite{OtterRoadMap}, which reports experimental results for several leading solvers across a variety data sets.  By any measure, however, the Eirene library outperforms existing packages by orders of magnitude in both time and memory on certain classes of interest \cite{henselman2017}, and provides a greater level of detail concerning the relationship between inputs and barcodes  than is standard among persistence solvers \cite{henselmanghrist2016}.

Libraries of this type generally begin by constructing a filtered simplicial complex
from a point cloud and storing the information in a matrix.
They next apply matrix reduction techniques and obtains the 
intervals of the barcode by pairing the simplices in the reduced matrix.
As mentioned before, the key feature of Eirene is that it adopts discrete
Morse Theory to reduce the matrix size and uses the Schur complement 
to perform the matrix reduction efficiently. 

\subsection{Persistence Diagrams}   \label{sec:persistencediagrams}

\newcommand{\dparam}{\epsilon}
\newcommand{\im}{{\mathrm{Im}}}
\newcommand{\field}{\mathbb K}
\newcommand{\spann}{\mathrm{span}}
\newcommand{\low}{\mathrm{low}}
\renewcommand{\max}{\mathrm{max} \; }
\newcommand{\boundary}{D}
\newcommand{\R}{\mathbb{R}}
\renewcommand{\partial}{D}

 The filtered simplicial complexes constructed to model scientific data are frequently large,  with billions and even trillions of component cells.  The standard method to compute the persistent homology of a filtered complex $X$ involves a large number of operations on the boundary operator of $X$.  The boundary operator is a matrix $D$ whose rows and columns are labeled by the simplices of $X$, and whose entries are determined by a mathematical formula.  Constructing and storing boundary matrices for large spaces is computationally expensive.  This is the central challenge to high-performance persistent homology solvers.  

State of the art platforms circumvent the problem by computing persistence diagrams (equivalently, barcodes) without constructing the full boundary matrix $\boundary$ of $X$, via specialized algorithms, data structures, and mathematical optimizations.

\begin{algorithm}
\caption{Column Algorithm}\label{alg:column}
\begin{algorithmic}[1]
\Procedure{pHcol($D$)}{} \Comment{$D = D_{i = 1, \ldots, n}^{j = 1, \ldots, n}$ is an $n \times n$ matrix }
\State{$R \gets D$, $V \gets I$} 
\For{$i=1, \ldots, n$} 
	\While  {$\exists j < i$ with $\low_{R}(j) = \low_R(i)$}
                \State{$c = R[\low_R i, i] / R[\low_R j, j]$ }
                \State{$R[...,i] = R[...,i] - c R[...,j]$} 
                \State{$V[...,i] = V[...,i] - c R[...,j]$}
	\EndWhile
\EndFor
\State{$S \gets \{(\low_{R} i, i) : R[ ..., i] \neq 0\}$ } 
\label{euclidendwhile}
\State \textbf{return} $R,V,S$ \Comment{$R$ is the reduced matrix,  $V$ is the column transformation matrix, and $S$ is the set of pivots.}
\EndProcedure
\end{algorithmic}
\end{algorithm}

\begin{algorithm}
\caption{Row Algorithm}\label{alg:row}
\begin{algorithmic}[1]
\Procedure{pHrow($D$)}{} \Comment{$D = D_{i = 1, \ldots, n}^{j = 1, \ldots, n}$ is an $n \times n$ matrix }
\State{$R \gets D$, $V \gets I$} 
\For{$i=n$ \textbf{down to} $1$} 
	\State{indices = $[j | \low_R(j) = i]$}
	\State{$p$ = indices$[0]$}	
	\For  {$j \in$ {indices}$[1..]$}
                \State{$c = R[\low_R i, i] / R[\low_R j, j]$ }
                \State{$R[...,i] = R[...,i] - c R[...,j]$} 
                \State{$V[...,i] = V[...,i] - c R[...,j]$}
	\EndFor
\EndFor
\State{$S \gets \{(\low_{R} i, i) : R[ ..., i] \neq 0\}$ } 
\label{euclidendwhile}
\State \textbf{return} $R,V,S$ \Comment{$R$ is the reduced matrix,  $V$ is the column transformation matrix, and $S$ is the set of pivots.}
\EndProcedure
\end{algorithmic}
\end{algorithm}

\subsubsection{Algorithms} \label{subsec:algorithms}
The primary mechanism of most solvers is a matrix decomposition $\boundary V = R$, where $V$ is invertible, upper-triangular, and block-diagonal (blocks correspond to groupings of simplices according to dimension), and $R$ is \emph{reduced}, in the following sense \cite{CEMVines06,SMVDualities11}.  For any matrix $A$, define $\low_A(j) = \max\{i : A[i,j] \neq 0\}$.  To wit, $\low_A(j)$ is the index of the lowest nonzero entry in column $j$.  This function is not defined on zero columns.  A matrix $R$ is \emph{reduced} if $\low_R$ is injective on its domain of definition. If $t_j$ is the time at which simplex $j$ enters the filtration, if every simplex enters at a different time, and if the row and column labels of the boundary matrix $\boundary$ are ordered according to $t$, then for any decomposition $\boundary V = R$ the set 
\begin{align*}
\textsc{Pers}(X) = \{ (t_j,t_{\low_R(j)})  : \low_R \textrm{ is defined on $j$}\}
\end{align*}
is the persistence diagram of the filtered complex.   Here $j$ runs over all column-indices in the domain of $\low_R$.

Note that the elements of this set are ordered pairs of real numbers.  If one regards $(t_j,t_{\low_R(j)})$ not as an ordered pair but as an interval on the real number line, then one obtains the barcode of $X$:
\begin{align*}
\textsc{Barcode}(X) = \{ [t_j,t_{\low_R(j)}] \subseteq \R : \low_R \textrm{ is defined on $j$}\}
\end{align*}

  Two standard algorithms to compute $V$ and $R$ are Algorithm \ref{alg:column}, also called the the \emph{column} algorithm, and Algorithm \ref{alg:row}, also called the \emph{row} algorithm \cite{SMVDualities11}. 

\begin{figure}
\centerline{}
\centerline{\includegraphics[width=4.25in]{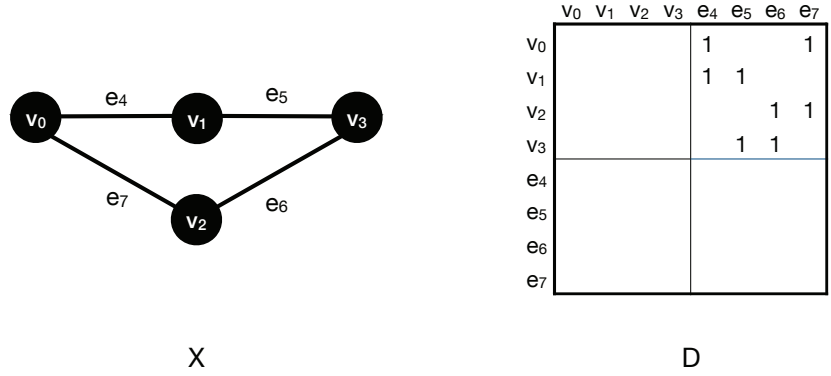}}
\caption{\textit{Left} A 1-dimensional simplicial complex.  Every 1-dimensional simplicial complex is a combinatorial graph.  We define a filtration on $X$ by $X_{t} = \{ v_s : s \le t\} \cup \{ e_s : s \le t\}$.  See, for example, Figure \ref{algex_barcode}.  \textit{Right} The  boundary matrix of $X$.}
\label{algex_complexmatrix}
\centerline{}
\centerline{}
\end{figure}

\begin{figure}[h]
\centerline{}
\centerline{\includegraphics[width=6.5in]{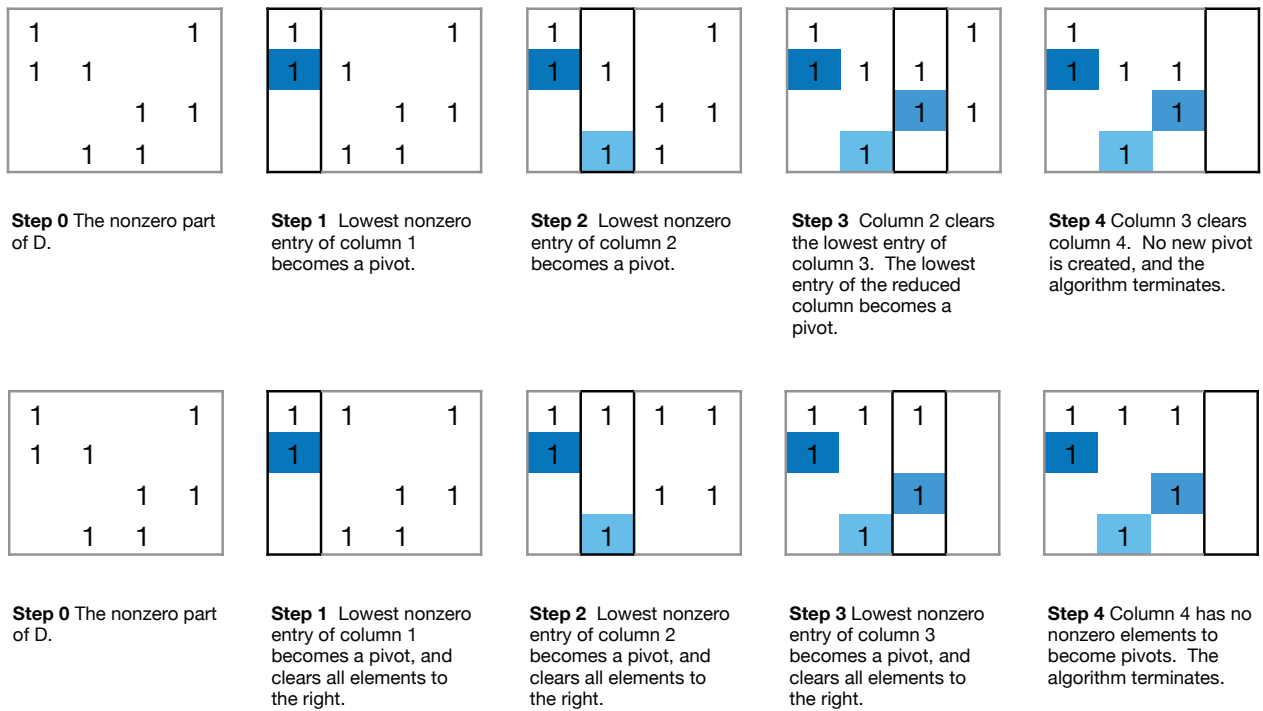}}
\caption{\textit{Top row} An application of the Column Algorithm to  the boundary matrix $D$ in Figure \ref{algex_complexmatrix}.  Only the nonzero part of $D$ is shown. \textit{Bottom row} An application of the Row Algorithm to the the boundary matrix $D$ in Figure \ref{algex_complexmatrix}.  Only the nonzero part of $D$ is shown.}
\label{algex_barcode}
\centerline{}
\centerline{}
\end{figure}

\begin{figure}[h]
\centerline{}
\centerline{\includegraphics[width=5.75in]{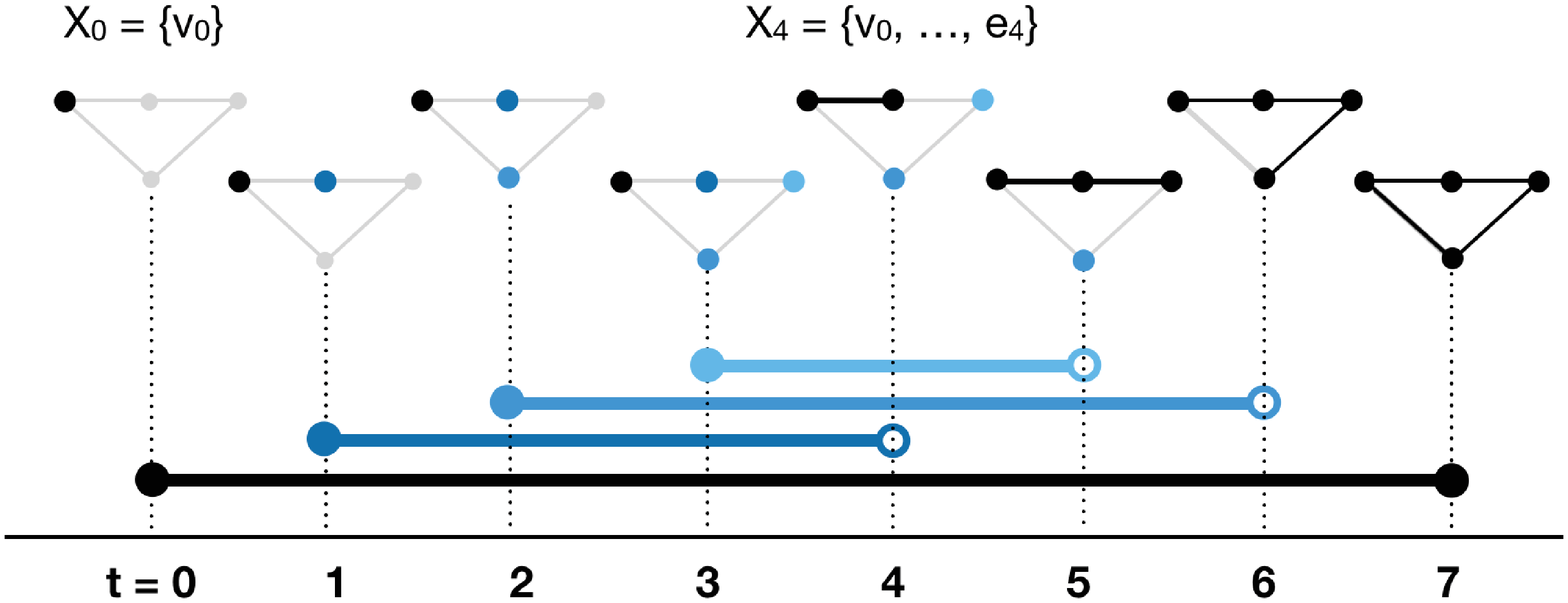}}
\caption{ The dimension-0 barcode of the filtered complex $X_{t} = \{ v_s : s \le t\} \cup \{ e_s : s \le t\}$ presented in Figure \ref{algex_complexmatrix}.  Dimension-0 homology counts the number of connected components of a space, so the number of bars at time $t$ equals the number of connected components at time $t$.   There is one blue bar for each pivot element shown in Figure \ref{algex_gauss} (equivalently, for each pivot element in Figure \ref{algex_barcode}).  The birth and death time for each bar may be read from the corresponding pivot.  For example, the light blue bar with endpoints $(3,5)$ corresponds to pivot  $(v_3, e_5)$, located in row 4, column 2 of the reduced boundary matrix.   In dimension 0 only, one additional  bar must be added, with birth time 0 and death time $\infty$.  This is shown in black.  }
\label{algex_barcode}
\centerline{}
\centerline{}
\end{figure}

\subsubsection{Optimizations and applications of the reported improvements} \label{subsubsec:barcodeoptimizations}

Several optimizations are available for both Algorithms \ref{alg:column} and \ref{alg:row}.  Each of those described below offers significant  improvements in time and memory.  Those employed by Eirene are specially marked.

\noindent {General optimizations}

\begin{itemize}
\item[]\textbf{Morse reduction} (Eirene) A preprocessing step motivated by smooth geometry, which reduces the size of $\partial$ before it is passed to Algorithm \ref{alg:column} or \ref{alg:row}.  This is one of the principle methods employed by the software package \textsc{Perseus} \cite{Perseus}.   Eirene applies a specialized form of Morse reduction which has been observed empirically to reduce the size of the boundary operator by multiple orders of magnitude \cite{HenselmanThesis}.
\item[]\textbf{Clear and compress} (Eirene) A preprocessing step motivated by homological algebra, which   reduces the size of $\partial$ before it is passed to either \textsc{Pers} or \textsc{LU}.  This method is employed in the platform PHAT \cite{Bauer2014, BKRClear14}.
 \item[] \textbf{Data structures}    Algorithms \ref{alg:column} and \ref{alg:row} depend heavily on efficient look-up of the lowest nonzero entry in each column.  Moreover, the columns of $\boundary$ that are eventually cleared play a very minor role.  For this reason, it has been found advantageous to store boundary matrices in compressed formats which can be queried for specific column/row information.  PHAT, for example, stores sparse columns as \emph{heaps}, which facilitate efficient queries to $\low_R$.  Simplex trees and combinatorial number systems have likewise been applied to reduce the cost of storing and manipulating the boundary matrix, e.g.\ in GUDHI.  

 \item[] \textbf{(Re)construction on the fly} On the extreme end of compression schemes, it is possible to \emph{avoid} the construction of $\partial$ altogether.  In this strategy, one does not construct column $j$ of the boundary matrix $\partial$ until iteration $j$.  One then reconstructs the preceding pivot columns as needed to reduce $j$.  This method is implemented in the software platform Ripser, and has achieved remarkable performance improvements.
 \end{itemize}
 
 \noindent {Optimizations for persistence diagrams only}
 
\begin{itemize}
\item[]\textbf{Persistent cohomology} If one  wishes to compute persistence diagrams only, and no cycle representatives, then there exists a  powerful optimization called persistent cohomology.  This method was enabled by a mathematical insight of Cohen-Steiner et al.\ \cite{CEMVines06}, and was concretely described and implemented by Morozov et al.\ \cite{SMVDualities11}.   These researchers showed that Algorithms \ref{alg:column} and \ref{alg:row} will return the proper pivot elements for a matrix $M$ if they are passed the antitranspose of $M$.  For most scientific data sets, it has been observed empirically that (when paired with clear and compress optimization) the time and memory performance of both algorithms improves by several orders magnitude when the anti-transpose is passed, rather than the original input.  We are not aware of any theoretical guarantees for this performance increase, as asymptotic behavior on scientific data is extraordinarily difficult to analyze.   The phenomenon is very robust in practice, however.  See for example \cite{SMVDualities11}.

\end{itemize}

  \noindent Optimizations special to Algorithm \ref{alg:row}
  
  \begin{remark}  These optimizations are used by Eirene, but  they can be applied to any implementation of Algorithm  \ref{alg:row}.
\end{remark}

  \begin{enumerate}
  \item[]\textbf{Block Reduction} (Eirene) This method applies the mathematics of algebraic \emph{Morse theory} to consolidate the clearing operations of many pivots simultaneously -- in effect, block clearing operations.  The matrices produced by such operations can be expressed in terms of Schur complements.  The calculation of Schur complements involves matrix inversion, multiplication, and addition -- these are the operations that are improved by the optimization described in \S\ref{subsec:sparsmatmul} below.
  \item[]\textbf{Dynamic reordering of rows and columns} (Eirene) The birth time of simplices does not determine a total ordering on the rows and columns of $\partial$ in general;\ this is because many simplices in the same complex may have identical birth times.  It has been observed that the choice of row and column order dramatically impacts matrix fill from clearing operations \cite{HenselmanThesis}.  Moreover, improved row and column orders may be deduced from the structure of the boundary matrix as an increasing number of operations are performed.  Thus Eirene dynamically reorders rows and columns following each column reduction.  These reordering operations were improved by the optimization of sorting functions described in \S\ref{subsec:redundant}.
\item[]\textbf{Specialized Vector Fields}  (Eirene) A critical component of the algorithms employed by Eirene to reorder rows and columns is the calculation of Morse vector fields.  These are mathematical objects which, in the current context, are entirely determined by the linear order of rows and columns in the distance matrix $M$.  Strategic reordering of the rows/columns of $M$ has been shown to produce much improved vector fields, and the calculation of ``good'' permutations on $M$ is a second major application of the improved sorting algorithms discussed in  \S\ref{subsec:redundant}.
  \end{enumerate}

\subsection{Persistent Generators: The Novel Contribution of Eirene}   \label{subsec:generators}

A set of {generators} in persistent homology is a collection of cycle representatives (formally, linear combinations of simplices) that parametrize or ``generate'' the bars of the barcode.  Generators in persistent homology may be obtained in several ways.   Cohen-Steiner et al.\ \cite{CEMVines06} showed that generators may be obtained from the columns of the matrix $V$ in the matrix  decomposition $D = RV$.

A fundamental obstacle to efficient generator calculation is the sparcity structure of $V$.  In scientific applications the superdiagonal blocks of $\partial$ are short and wide, with columns outnumbering rows by several orders of magnitude.  This means that at least one nonzero block of  matrix $V$ is large -- so large as to be impractical to store or manipulate, even for mesoscale data sets.    

The \emph{cohomology algorithm} circumvents this problem by sacrificing the ability to obtain generators.  It applies Algorithm \ref{alg:row} to the antitranpose $\partial^\perp$.  The $V$ of this decomposition has much smaller blocks, on the whole, and therefore vastly less scope for fill.  However, the $V$ from this decomposition does not provide generators.

With the exception of Eirene, every  state of the art persistence solver relies on an variant of the cohomology algorithm to obtain competitive  performance results.  The time and memory cost to compute generators on cohomology-based platforms is generally several orders of magnitude greater than that needed to compute barcodes.   At the current time, Eirene is the only platform capable of computing homological generators within an order of magnitude of the best recorded cost to compute barcodes.  This is important, as the computation of generators is \emph{fundamentally different} to, and strictly harder than, that of barcodes.   

A survey of available platforms and run times for barcode computation -- without generators -- was recently released in \cite{OtterRoadMap}.   The primary findings of this survey and comparison run times of Eirene are available in \cite{HenselmanThesis}.  The fundamental insight of Eirene is a combinatorial, matroid-theoretic interpretation of persistence calculation which (a) admits calculation of generators by elementary matrix operations, and (b) reduces fill in the matrix $\partial$ during reduction \cite{HenselmanThesis}.

\section{WorldMap: an Example of using Eirene}

\begin{figure}
\centerline{\includegraphics[width=3.5in]{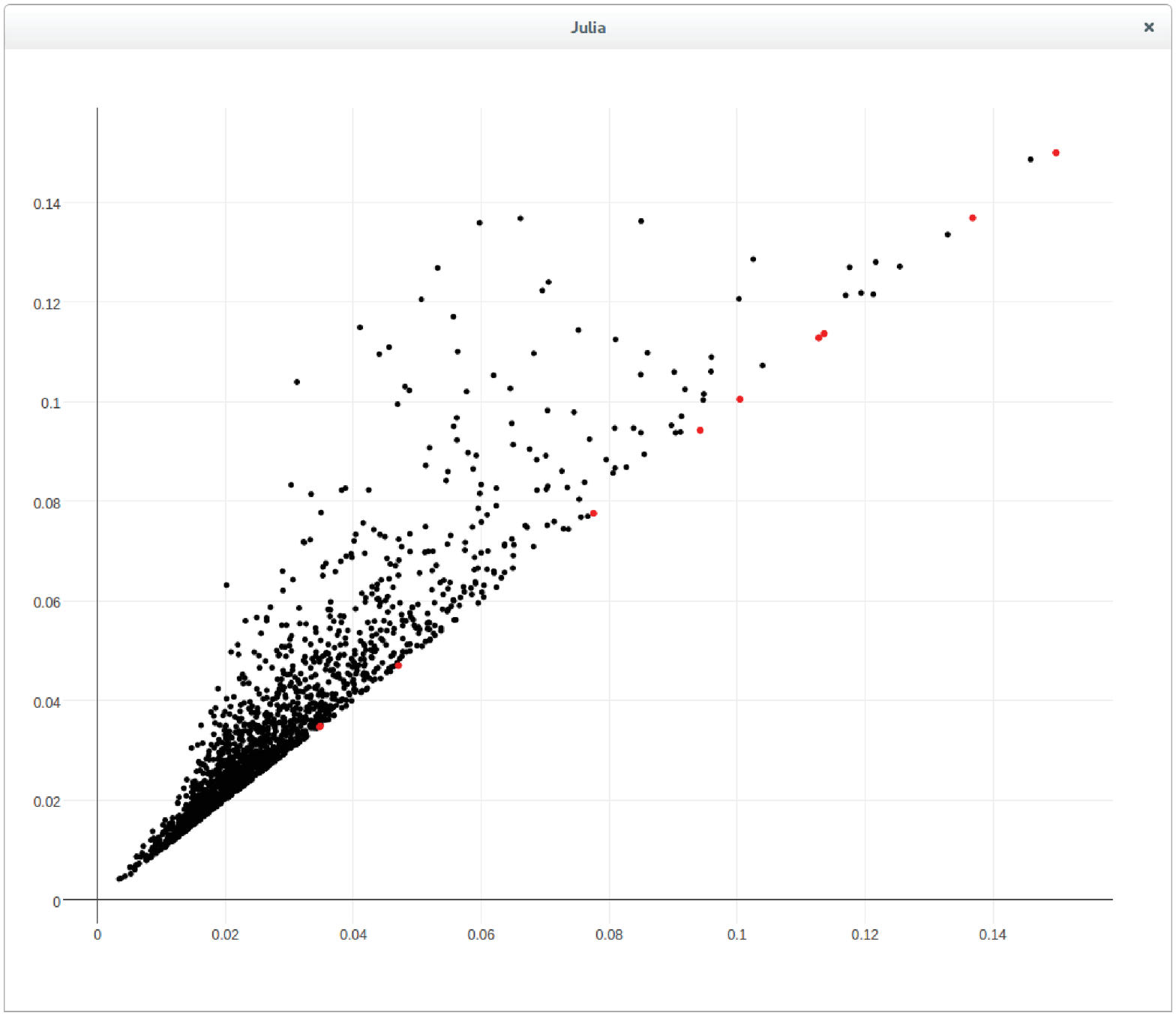}}
\caption{The Persistent Diagram of the WorldMap ($\epsilon=0.15$)}
\label{PersistentDiagram}

\centerline{}
\centerline{}
\centerline{}
\centerline{\includegraphics[width=3.75in]{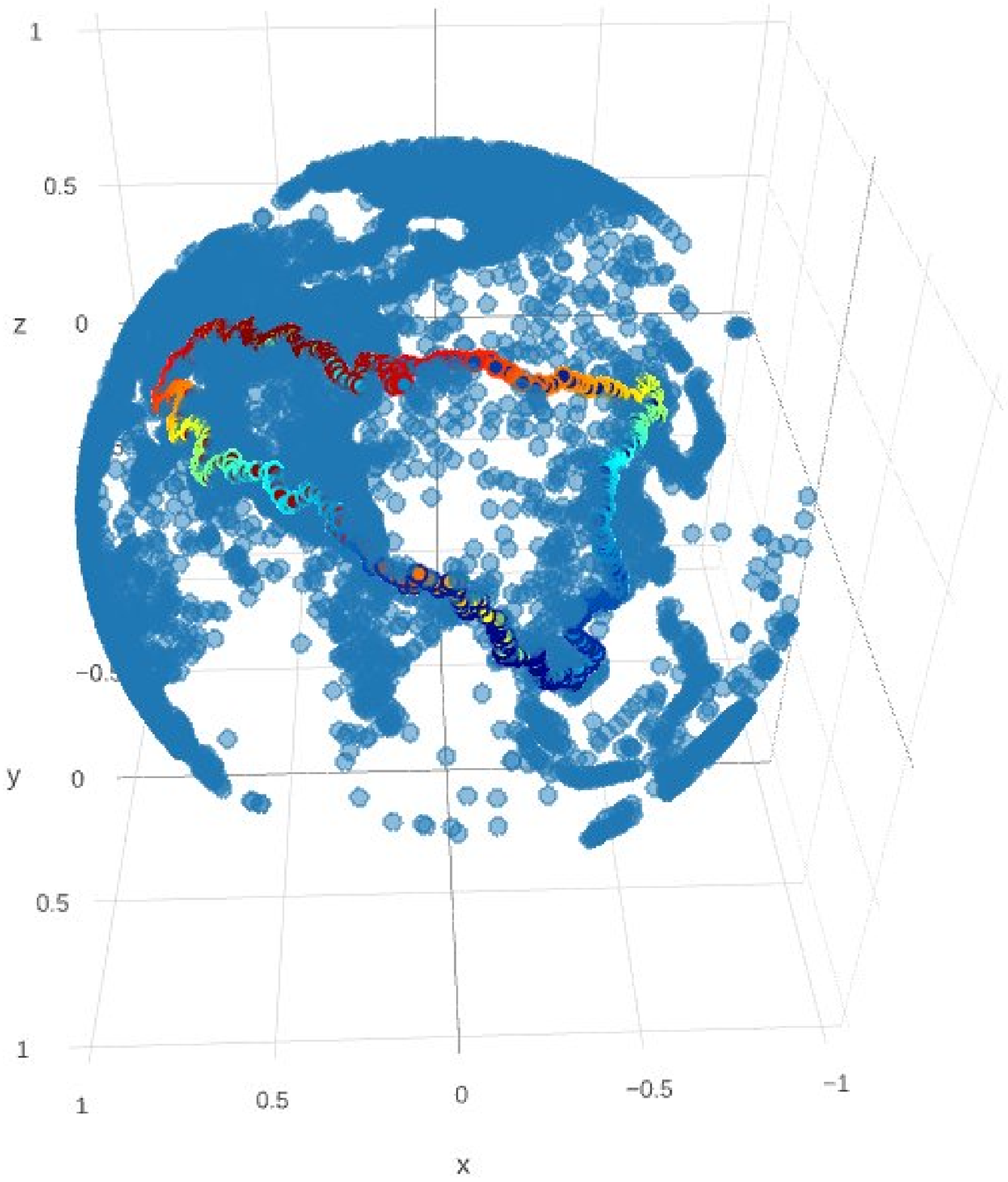}}
\caption{A Cycle in the WorldMap displayed by the Eirene tool}
\label{EireneWorldMap}
\end{figure}

In this section, we demonstrate an example of how to use Eirene to load data,
compute the persistent homology classes, and plot the results graphically.  The basic syntax of Eirene is a call to the {\tt ezread()} function:
\begin{Verbatim}
     julia> C = eirene(x, keywords ...)
\end{Verbatim}
where $x$ is either a point cloud encoded as the rows or columns of an array, or a square symmetric matrix (typically a pairwise distance matrix).  Keyword arguments specify how this input matrix should be interpreted and processed, and what topological features should be extracted (e.g.\ homology groups in the first 5 dimensions).  The package computes this information and stores the output in a dictionary object {\tt C}.  This object can then be queried for specific information, e.g.\ Betti curves, or passed to specialized functions for 3D plotting, c.f.\ Figures \ref{PersistentDiagram} and \ref{EireneWorldMap}.

For a concrete example, suppose we want to explore world geography vis-a-vis networks of neighboring cities.  
There is a wealth of data available online, and for this demo example we use a 
catalog of 7322 cities which can be found on the web site {\tt simplemaps.com}. 
After downloading the {\tt .csv} data file, start running a Julia 
REPL(Read-Evaluate-Print-Loop) to include the Eirene code and then load the data 
into a 2-dimensional array using the Eirene {\tt ezread()} wrapper, as shown below.

\begingroup
\scriptsize
\begin{Verbatim}
     julia> include("<file_path_to_Eirene>") 
     julia> a = ezread("<file_path_to_csv_file>")
     7323x9 Array{Any,2}:
     "city"          "city_ascii"    "lat"   "lng" ... "country"     "iso2" "iso3" "province"       
     "Qal eh-ye Now" "Qal eh-ye"     34.983  63.1333   "Afghanistan" "AF"   "AFG"  "Badghis"         
     "Chaghcharan"   "Chaghcharan"   34.5167 65.25     "Afghanistan" "AF"   "AFG"  "Ghor"    
     "Lashkar Gah"   "Lashkar Gah"   31.583  64.36     "Afghanistan" "AF"   "AFG"  "Hilmand"         
     "Zaranj"        "Zaranj"        31.112  61.887    "Afghanistan" "AF"   "AFG"  "Nimroz"          
     ...
\end{Verbatim}
\endgroup

Then, we can extract the spherical coordinates, i.e., columns 3 and 4, from the 
array {\tt a} and invoke the function {\tt convert()} to convert the data type 
from Any to {\tt Float64}. 
Note that Eirene has a built-in function {\tt latlon2euc()}  which can be used to
translate spherical coordinates (in degrees, with fixed radius 1) 
to 3D Euclidean coordinates. The {\tt rowsare} keyword argument determines wether 
rows are treated as points or dimensions.

\begingroup
\scriptsize
\begin{Verbatim}
     julia> b = a[2:end,3:4]
     7322x2 Array{Any,2}:
       34.983   63.1333
       34.5167  65.25  
       31.583   64.36  
       ...
     julia> b = convert(Array{Float64,2},b)
     7322x2 Array{Float64,2}:
       34.983   63.1333
       34.5167  65.25  
       31.583   64.36  
       ...
     julia> c = latlon2euc(b,rowsare="points")
     3x7322 Array{Float64,2}:
     0.370265  0.344959  0.368622  0.403432  0.344318  ...  0.81567    0.824296
     0.730885  0.748275  0.767998  0.755149  0.768533  ...  0.490971   0.449048
     0.573333  0.566646  0.523733  0.516713  0.53926   ... -0.305991  -0.344807
\end{Verbatim}
\endgroup

To pass city names to the Eirene function, we need to extract 
the column 2 of array {\tt a}.
Due to potential errors resulting from nonstandard character strings, it's good practice to use 
the built-in label sanitzer {\tt ezlabel()} to clean this column before assigning to the
array {\tt d}.  The wrapper will 
replace any element of the column that cannot be expressed as an ASCII String with the number 
corresponding to its row.  
\begingroup
\scriptsize
\begin{Verbatim}
     julia> d = ezlabel(a[2:end,2])
     7322-element Array{Any,1}:
     "Qal eh-ye"  
     "Chaghcharan"
     "Lashkar Gah"
     ...
\end{Verbatim}
\endgroup
With our inputs in order, it's time to call the main function {\tt eirene()}. 
A cursory inspection shows that a number of interesting features appear at or below 
distance threshold $\epsilon  = 0.15$ (i.e. approximate to 0.15 * earth radius = 956Km), 
so we'll use that as our initial cutoff.  As always, the first argument is the most important: this is the matrix {\tt c}, from which the software will build a reduced Vietoris-Rips complex (this construction is described in \S\ref{subsec:homologicalpersistence}).  Keyword {\tt rowsare = ``dimensions''} declares that the rows of {\tt c} should be regarded as dimensions in Euclidean space, and, conversely, that the columns of {\tt c} should be regarded as the vertices or 0-simplices in the Vietoris-Rips construction.  The properties of this complex computed by Eirene will be stored in a dictionary object $C$, which can be queried with various specialized functions.

\begingroup
\scriptsize
\begin{Verbatim}
     julia> C = eirene(c,rowsare = "dimensions",upperlim = 0.15,pointlabels=d)
     elapsed time: 74.892521378 seconds
     Dict{ASCIIString,Any} with 14 entries:
       "symmat"               => 7322x7322 Array{Int64,2}:
       "filtration"           => Any[[515055,515055,515055,515055,515055 ...
       ...
\end{Verbatim}
\endgroup

Because we did not specify the {\tt bettimax} parameter when calling the {\tt eirene()} function,
the default is to compute the persistence modules in dimensions 0 and 1. 
We can use the function {\tt plotpersistencediagram\_pjs()} to view the persistence diagram,
as shown in Figure~\ref{PersistentDiagram}. Note that the persistence diagram is an
alternative graphical way to represent barcodes. That is, the x-coordinate and the y-coordinate
in a persistence diagram represent the birth and death times, respectively. 
Bars of infinite length appear in red at the point on the diagonal corresponding to the time
of their birth.  Hovering over a point in the diagram will display the identification number 
of the corresponding persistent homology class, together with the size the cycle 
representative Eirene computed for it.  To view a specific cycle, we can invoke the 
function {\tt plotclassrep\_pjs()} by passing the identification number of the class
as a parameter.

\begingroup
\scriptsize
\begin{Verbatim}
     julia> plotpersistencediagram_pjs(C)
     julia> plotclassrep_pjs(C,class = 1757)
\end{Verbatim}
\endgroup

Figure~\ref{EireneWorldMap} is a 3D visualization of the persistent 1-cycle with the class id
number 1757, generated from WorldMap data regarding cities embedded in the Eurasian continent: 
the Himalayan branch of the silk road is clearly 
visible on its South West arc; to the North it follows the Trans-Siberian railroad from Moscow in 
the west to Vladivostok on the Sea of Japan, by way of Omsk, Irktsk, and Chita; from there, it 
follows the connecting route from Beijing to Hong Kong, passing Nanning, Hanoi, and close to 
Ho Chi Min on its way to Bangkok (not all of these appear in the cycle itself, but they are 
easily spotted nearby).  With a little exploration we can find large features formed by the 
Sahara Desert, Tapajos River, Falkland Islands, South China Sea, Guam, and Hudson Bay, and 
smaller ones shaped by the Andes mountains and Gulf of Mexico - all through the proxy of 
urban development.

\section{Performance-Improving Methods}

Code analysis tools are important for programmers to understand program behavior. 
Software profiling measures the time and memory used during the execution of a 
program to gain this understanding and thus helps in optimizing  code.
To develop efficient software, it is essential to identify the major
bottlenecks and focus optimization efforts on these.
Therefore, our strategy is to use Julia's built-in
Profile module, a statistical profiler, 
to find the key bottlenecks in Eirene and  develop
different performance-improving methods to solve them. 

To profile an execution of Eirene, we simply need to put the macro {\tt @profile}
before calling the main function of Eirene, e.g. \\
\mbox{\hskip 0.15 in \scriptsize \tt @profile C = eirene(data-file-path,...) } \\ 
Note that the Julia profiling tool works by periodically taking a backtrace 
during the execution of a program. Each backtrace takes a snapshot of the 
current state of execution, i.e. the current running function and line number along
with the complete chain of function calls which led to this line. 
Therefore, a busy line of code, such as the code inside nested loops,
has a higher likelihood to be sampled and hence appears more frequently 
in the set of all backtraces. However, profiling a very long-running task  
may cause the backtrace buffer to fill. Programmers can use the configuration
function {\tt  Profile.init(n, delay)} to either increase the total number 
of backtrace instruction pointers {\tt n} (default: $10^6$) or 
the sampling interval {\tt delay} (default: $10^{-3}$) or both.  
    
We used three benchmarks: HIV, WorldMap, and Dragon2, to locate the 
bottlenecks in Eirene. The HIV benchmark contains the Hamming distances
between 1088 different genomic sequence of the HIV virus. The WorldMap
benchmark includes the data of 7322 cities in the world. The Dragon2
data set contains 2000 points sampled from the Stanford Dragon graphic. 
Both of the HIV and the Dragon2 benchmarks are available from \cite{OtterRoadMap}, while
the WorldMap data can be obtained from \cite{Eirene}.  

After the major bottlenecks were identified, we figured out how each was formed and developed targeted solutions,
as described below. 

\subsection{Organization}

Recall from \S\ref{sec:persistencediagrams} and \S\ref{subsec:generators} that Eirene executes a variant of Algorithm \ref{alg:row}.  Rather than performing row-clearing operations one at a time, this variant eliminates multiple rows in a single pass, via block clearing operations.    The primary input to the software is an $m \times m$ distance matrix $S$.  The distance matrix mathematically determines an $n \times n$ filtered boundary operator $D$.  In general, $n$ is many orders of magnitude larger than $m$.   Pseudocode for the work flow appears in Algorithm \ref{alg:morse}.
\begin{algorithm}
\caption{Morse Reduction}\label{alg:morse}
\begin{algorithmic}[1]
\Procedure{Eirene($S$)}{} \Comment{$S = S_{i = 1, \ldots, m}^{j = 1, \ldots, m}$ is an $m \times m$ distance matrix, with boundary operator $D$ }
\State{Directly construct a submatrix $E$ of the boundary operator $D$.  Matrix $E$ contains all data needed to complete the calculation.  $D$ is never stored in memory.}
\While{$E^T$ has unreduced columns} 
                \State{Identify a filtration-compatible discrete Morse vector field $M$ (a special type of square, invertible submatrix).}
                \State{As in Algorithm \ref{alg:row}, perform a block pivot on $M$ to clear all entries to the right in $E^T$.} 
\EndWhile
\label{euclidendwhile}
\State{ \textbf{return} the set of pivots and the row/column operation matrices that realize this reduction.  This suffices to calculate a complete set of generators for the filtered complex. }
\EndProcedure
\end{algorithmic}
\end{algorithm}

Our performance-improving methods address five central issues in this work flow: (a) arrays of type Any, (b) redundant calculations, (c) deeply nested loops, (d) sorting large arrays, and (e) sparse matrix multiplication and addition.  These improvements impact the following  stages of the work flow.

\begin{enumerate}

\item[] \textbf{Construction of $E$}: {(a) arrays of type Any,  (b) redundant calculations, (c) deeply nested loops}\\
Efficient construction of $E$ depends on the reordering of large lists of simplices according to various conditions on their  geometric properties.  We improved performance by modifying the declared types of some arrays, which were employed at this stage, and by eliminating redundant calculations and deeply nested loops.   

\item[] \textbf{Construction of Morse vector fields}: (d) sorting large arrays\\
 Eirene applies a principled procedure to select vector fields with advantageous properties at each iteration of the while loop.  This procedure has been demonstrated empirically to reduce fill in the submatrix $E$.  This method requires reordering of the rows and columns of $E$.  We improved the performance by implementing a different sorting algorithm on GPU.
 
\item[] \textbf{Block clearing}: (e) sparse matrix multiplication and addition\\
The block clearing operation performed on each iteration of the while loop in Algorithm \ref{alg:morse} is functionally equivalent to computation of the Schur complement of $M$ in $E$.   We parallelized this computation with a master/workers model.
\end{enumerate}

The remainder of this section provides a detailed discussion of points (a)-(e).  

 \begin{remark} Every persistent homology solver that operates on the principle of discrete Morse theory follows the work flow laid out in Algorithm \ref{alg:morse}.  Such implementations will necessarily involve similar variable types and operations: permutations determined by geometric features, arrays of arrays, etc.    Consequently, improvements  (a)-(e) will adapt directly to any such platform.  For example, they will apply to any Eirene-like solver implemented in python or C++.
 \end{remark}

\subsection{Arrays of type Any}

Julia's sampling profiler  displays  results in
textual format only. For faster comprehension,
we used the ProfileView package's function {\tt ProfilevView.view()}. 
This function plots a visual representation of the call graph called 
a flame graph~\cite{BGregg}.  
The vertical axis of the flame graph (from bottom to top)
represents the stack of function calls, 
while the horizontal axis represents the number of backtraces  sampled
at each line.  To identify a potential bottleneck, the user can hover a cursor over a
long bar, usually  on the top two levels in the graph, and the corresponding     
backtrace of the function name and the line number will be shown on the screen. 

\begin{figure}
\centerline{\includegraphics[width=3.5in]{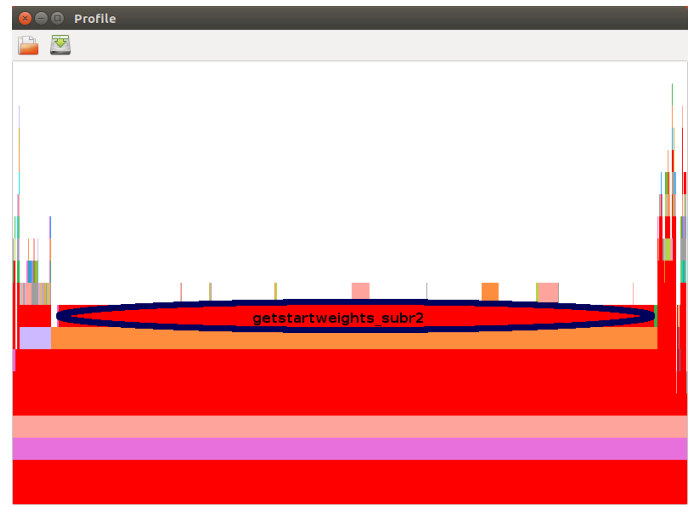}}
\caption{Identification of the bottleneck caused by the array of type Any }
\label{TypeANY}
\centerline{}
\end{figure}

Figure~\ref{TypeANY} shows the first performance bottleneck  identified, 
located in the function {\tt getstartweights\_subr2()} using the HIV benchmark.
It spans 90\% of the horizontal axis and similar results can be found in
the other two benchmarks. Examining the code for this function,
we found that the use of the array of the abstract-type {\tt Any} 
in the following three lines

\begingroup
\scriptsize
\begin{Verbatim}

     val = Array(Any,m)
     supp = Array(Any,m)
     suppDown = Array(Any,m)

\end{Verbatim}
\endgroup

\noindent causes the Julia interpreter to generate many dynamic invoker objects when
these three arrays appeared in an expression.  Performance
is greatly suffered as a result. Fortunately, the abstract-type {\tt Any} is unnecessary in this function because
the array {\tt supp[]} stores only indices (i.e. of integer type) 
of non-zero elements returned from the Julia function {\tt find()}. 
After replacing the abstract-type {\tt Any} with the static-type {\tt Int64}, this
bottleneck has been solved.

\centerline{}

\begin{figure}
\centerline{}
\centerline{\includegraphics[width=3.5in]{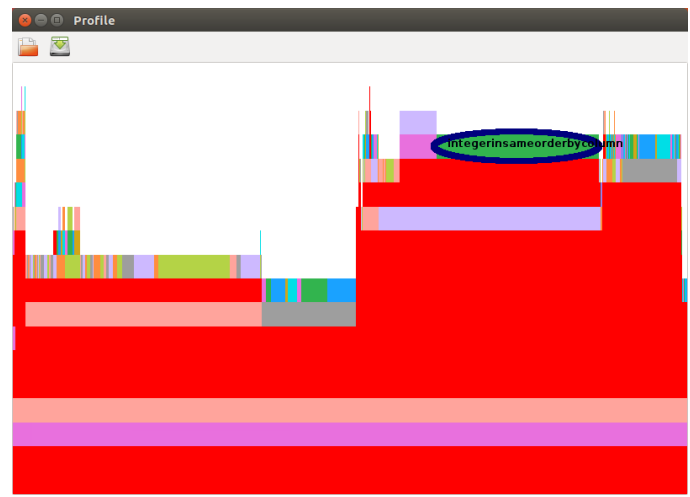}}
\caption{Identification of the bottleneck in the function integersinsameorderbycolumn}
\label{D2integersinsameorder}
\centerline{}
\centerline{}
\centerline{}
\end{figure}

\subsection{Redundant Calculations}
\label{subsec:redundant}

We ran the code and performed the profiling procedure again after fixing the first
large bottleneck. We noticed that there is a certain amount of time spent
in the function {\tt integersinsameorderbycolumn()} when running the Dragon2
benchmark, as shown in Figure~\ref{D2integersinsameorder}.  
Further investigation shows that this delay resulted from a number of unnecessary calculations. 
As shown in the upper box in Figure~\ref{JL_integersinsame}, the array {\tt y} is used to calculate 
the prefix sums of the array {\tt x}, only a fraction of which will be copied
to the array {\tt z} and returned to its caller. When the {\tt maxvalue}, 
a parameter passed to this function, is large, the inner loop
{\tt for i = 1:maxvalue} will let this function be executed much longer.   
As shown in the bottom box in Figure~\ref{JL_integersinsame},
we modified the code to find out the range first and then
calculate only the prefix sums within this range. 
Furthermore, we can just use the array {\tt x} to accumulate the prefix
sums of itself instead of using another local array {\tt y}.
The initialization of the whole array {\tt x} inside the beginning of the 
loop (i.e. {\tt x[:] = 0}) can also be replaced by cleaning up the dirty elements before
the end of the loop. Note that a single colon in Julia indicates every row or
column.

\begin{figure}
\centerline{}
\centerline{}
\scriptsize
\centering
\begin{Verbatim}[frame=single]
       ...
       for j = 1:numcols
                x[:] = 0
                for i = colptr[j]:(colptr[j+1]-1)
                        x[v[i]]+=1
                end
                y[1] = colptr[j]
                for i = 1:maxvalue
                        y[i+1]=y[i]+x[i]
                end
                for i = colptr[j]:(colptr[j+1]-1)
                        u = v[i]
                        z[i] = y[u]
                        y[u]+=1
                end
        end
        return z
\end{Verbatim}
\centerline{}
\centerline{}
\begin{Verbatim}[frame=single]
        ...
        x[:] = 0
        for j = 1:numcols
                for i = colptr[j]:(colptr[j+1]-1)
                        x[v[i]]+=1
                end

                maxv = v[colptr[j]];   minv = maxv
                for i = (colptr[j]+1):(colptr[j+1]-1)
                    if v[i] > maxv
                       maxv = v[i]
                    elseif v[i] < minv
                       minv = v[i]
                    end
                end

                prevsum = colptr[j]
                for i = minv:maxv
                        sum = prevsum + x[i]
                        x[i] = prevsum
                        prevsum = sum
                end
                for i = colptr[j]:(colptr[j+1]-1)
                        u = v[i]
                        z[i] = x[u]
                        x[u]+=1
                end

                for i = minv:maxv
                        x[i] = 0
                end
        end
        return z
\end{Verbatim}

\caption{The original(top) and the modified(bottom) function integersinsameorderbycolumn() in Eirene }
\label{JL_integersinsame}
\centerline{}
\centerline{}
\centerline{}
\end{figure}
 
\subsection{Deeply Nested Loops}

Even after we removed the overhead caused by the unnecessary
array of type {\tt Any}
in the function {\tt getstartweights\_subr2()}, the execution of 
the Dragon2 benchmark still spends a reasonable amount of time on
this function. As shown in Figure~\ref{JL_getstartwt}, this function calculates
the weight for each column in the matrix {\tt s}. It firstly
finds the indices of the non-zero elements in each column
and then uses nested loops to increment the counter of the 
column if certain conditions are met. Note that the array base function
{\tt find(X)} in Julia returns a vector of the linear indices of non-zeros in the array 
{\tt X}. Hence, the code {\tt find(s[:,i])} in Figure~\ref{JL_getstartwt}
will return the row indices of the non-zero 
elements in the {\tt i}th column.  When there are many
non-zero elements in the matrix {\tt s}, the deeply nested
loops will consume much CPU computation time. 

\begin{figure}
\centerline{}
\centerline{}
\scriptsize
\centering
\begin{Verbatim}[frame=single]
function getstartweights_subr2(s::Array{Int64,2},
            w::Array{Int64,1},m::Int64)
   ... 
   for i = 1:m
       supp[i] = find(s[:,i])
       l[i] = length(supp[i])
       ...
   end
   for i = 1:m
       Si = supp[i]
       ...
       for jp = 1:l[i]
           ...
           for ...  
              if conditions met 
                  w[i]+=1
              end
           end
       end
   end
end
\end{Verbatim}
\caption{The original function getstartweights\_subr2() in Eirene }
\label{JL_getstartwt}
\centerline{}
\centerline{}
\centerline{}
\end{figure}

To deal with the deeply nested loops, one feasible approach is to use the
GPU to accelerate the execution. NVIDIA provides a parallel computing platform
and programming model called CUDA (Compute Unified Device architecture).
Therefore, now it is much more convenient to write
application programs on the GPUs for processing large 
amounts of data, without the need to use low-level assembly language code.

The NVIDIA GPU architecture consists of a
scalable number of streaming multiprocessors (SMs),
each containing many streaming processors (SPs)
or cores to execute the lightweight threads.
The kernel function, which is declared by using the
{\tt \_\_global\_\_}  qualifier keyword in front of a function heading,
is executed on the GPU device.  It consists 
of a grid of threads and these threads are divided into a set
of blocks, each block containing multiple warps of
threads.  Blocks are distributed evenly to the different
SMs to run. A warp, which has 32 consecutive threads bundled together,
is executed using the Single Instruction, Multiple Threads (SIMT) style.
Note that the GPU device has its own off-chip device memory (i.e. global memory)
and on-chip faster memory such as registers and shared memory. 
Fancy warp shuffle functions are also supported in modern GPUs\cite{Harris}.
They permit exchanging of variables (i.e. registers) between
threads within a warp without using shared memory.

\begin{figure}
\centerline{}
\centerline{}
\centerline{\includegraphics[width=6in]{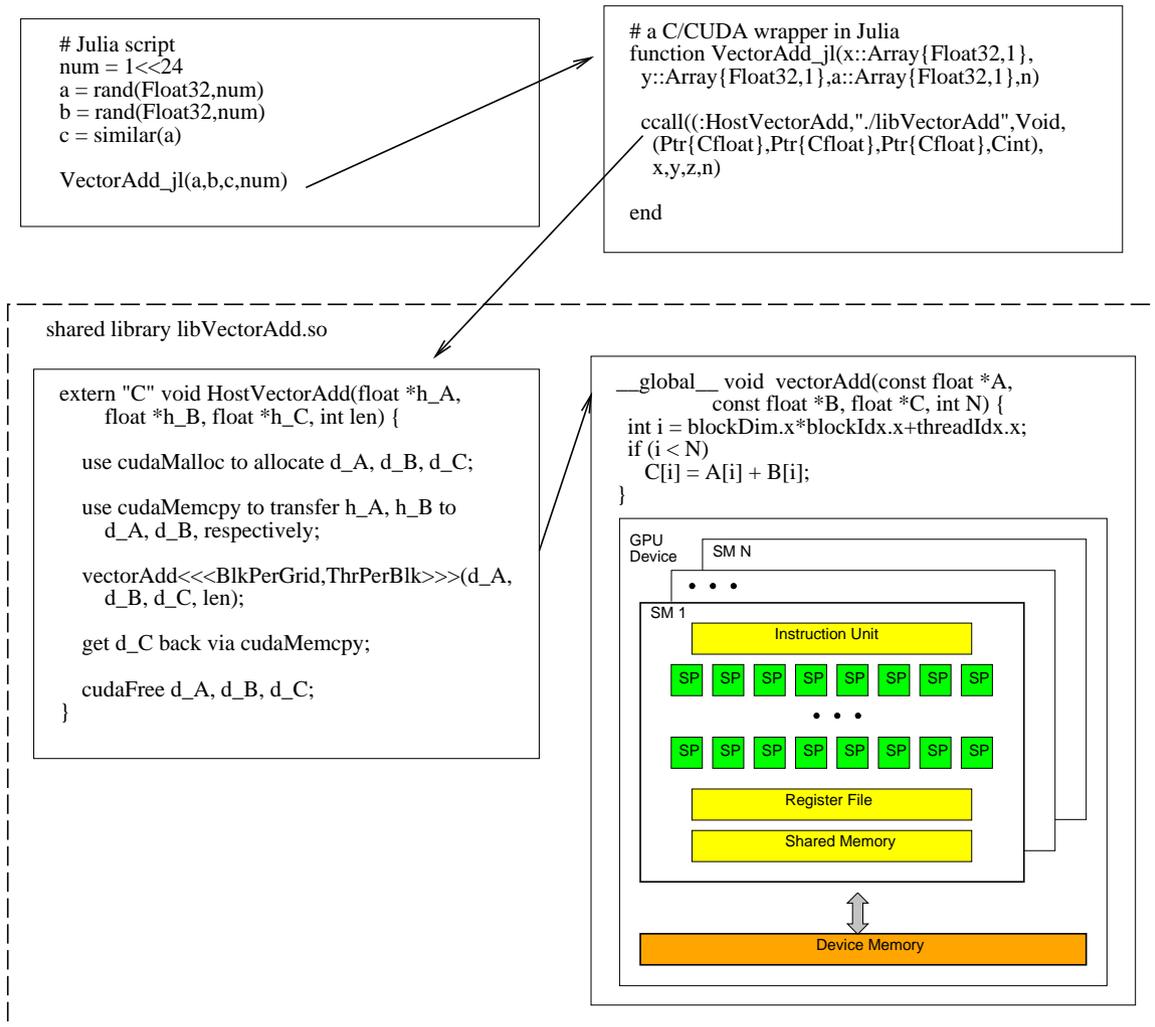}}
\caption{An example of calling C/CUDA functions from a Julia script}
\label{JuliaCcuda}
\centerline{}
\centerline{}
\centerline{}
\end{figure}

Though there are some Julia packages which enable programmers to launch
GPU kernel calls, we decided to implement our own wrappers
for greater flexibility and efficiency. Figure~\ref{JuliaCcuda}
shows an example of calling a CUDA function named {\tt vectorAdd()} by way
of the host function {\tt HostVectorAdd()} in C. A wrapper function in Julia
is needed, which utilizes the {\tt ccall()} function to invoke the host function.
Note that the first argument of {\tt ccall()} is a tuple pair
{\tt (:function,"library-path")}. The rest of the arguments include the function
return type, a tuple of input parameter types, and then the actual parameters.  
The C/CUDA functions should be compiled and linked as a shared objects.
It is worth mentioning that, when using the {\tt nvcc} NVIDIA CUDA Compiler
to compile the C/CUDA code, programmers need to use the options
{\tt  -Xcompiler -fPIC}  to pass the position-independent code (PIC) option 
from {\tt nvcc} to {\tt g++}.

\begin{figure}
\scriptsize
\centering
\begin{Verbatim}[frame=single]
_global__ void init_supp(const long long *s, int *supp, int *l, ..., long long m) {
   int tid = blockIdx.x * blockDim.x + threadIdx.x;
   int lnid = threadIdx.x % WARP_SIZE ; // lane id
   int warp_id = tid >> 5; // global warp number
   if(warp_id >= m) return;

   int supplen = 0;
   int j = lnid;

   while (j < m ) {
      int b = s[warp_id*m + j] != 0 ;
      int votes = __ballot( b ); // cast b if non-zero 
      int lidx = __popc( votes & ((1 << lnid) - 1)) ; 

      if (b)
        supp[warp_id*m + supplen+lidx] = j;
 
      supplen += __popc(votes);
      j += WARP_SIZE; // next stride
   }
   ...
   if(lnid == 0) {
      l[warp_id] = supplen;
       ...
   }
}

_global__ void calcstartweights(const long long *s, int *supp, int *l, long long *w, long long m) {
 // Same as in init_supp, calc. tid , lnid, and warp_id ; 
   if(warp_id >= m) return;

   int supplen = l[warp_id];
   int jp = lnid;
   int wt = 0; // each thread has a counter (in register) 
   while (jp < supplen) { 
      int j = supp[warp_id*m  + jp];
       ...
           for ... {
               if ( conditions met )
                 wt += 1;
           }
      jp += WARP_SIZE; // next stride
   }    
   // parallel reduction sum through registers
   for(int offset = WARP_SIZE>>1; offset>0; offset >>= 1) 
       wt += __shfl_down(wt, offset);

   if (lnid == 0) w[warp_id] = wt;
}

extern "C" void Host_getstarweights(long long * s, long long *h_w , long long m) {
   use cudaMalloc to allocate d_s, d_supp, d_l, d_w, etc.

   use cudaMemcpy to transfer data to d_s on GPU 

   init_supp<<<BlkPerGrid, ThrPerBlk>>>(d_s, d_supp, d_l,..., m);

   calcstartweights<<<BlkPerGrid, ThrPerBlk>>>(d_s, d_supp, d_l, d_w, m);

   use cudaMemcpy to get the weights h_w from device d_w 
}
\end{Verbatim}
\caption{Implementation of the getstartweights\_subr2() on GPU}
\label{GPU_getstartwt}
\end{figure}

Figure~\ref{GPU_getstartwt} shows our implementation of the function
{\tt getstartweights\_subr2()}. Our idea is to use {\tt m} warps
to handle the outermost {\tt for i=1:m} loop in the original code.  
Since the second {\tt for i=1:m} loop needs the result from the first loop,
we need to use two kernel functions:  {\tt init\_supp()} and
{\tt calcstartweights()}, and launch them one after the other
in the host function {\tt Host\_getstarweights()}. 
To find the indices
of the non-zero elements for each column, each thread within a warp 
checks the corresponding element and casts its one-bit vote via the
{\tt \_\_ballot()} intrinsic function. The {\tt \_\_ballot()}
collects the votes from all threads in a warp into a 32-bit integer and
returns this integer to every thread. The {\tt \_\_popc(int v)}
function returns the number of bits which are set to 1 in
the 32-bit integer v. That is, it performs the population count operation.
By combining the {\tt \_\_ballot()} and {\tt \_\_popc()} functions along with
bit-masking, each thread in a warp can quickly 
find out how many non-zero elements are in front of it, and then store
its index into the corresponding location in the array {\tt supp[]}.
This procedure will be repeated stride by stride until all
elements in a column have been processed.
Note that similar strategies have been used in \cite{Harris_Jade} and \cite{GCA16}
to perform efficient stream compactions on GPU. 
 
After determining the indices of the non-zero elements for each column,
each thread in the second kernel function {\tt calcstartweights()}  
uses a local counter (i.e. a register) and increments this counter by one
if certain conditions are met.  When all non-zero elements have
been checked, a parallel reduction sum operation via the efficient
shuffle function {\tt \_\_shfl\_down()} is performed. 
All of the local counter values in each warp will be added together and
stored into the counter at the first thread (i.e lane ID 0). This thread 
then writes the weight to the output array {\tt w}.

\subsection{Sortperm a Large Array}

Another bottleneck occurs in the function {\tt ordercanonicalform()}
when it calls Julia's {\tt steeper(v)} to find the rank of each element
in the distance matrix. The {\tt sortperm(v)} computes a permutation of the
array v's indices that puts the array into sorted order. For example,
if the input array {\tt v} is\\
\mbox{\tt \hskip 0.25in v = [ 7, 3, 8, 4, 2 ] }, \\
then the output from {\tt sortperm(v)} will be \\
\mbox{\tt \hskip 0.25in [ 5, 2, 4, 1, 3 ] }. \\
When the size of the 
matrix is large, e.g. the WorldMap benchmark, {\tt sortperm(v)} performs
poorly. We found out that if we change the default sorting algorithm
from MergeSort to RadixSort, the performance can be greatly improved,
especially for the WorldMap benchmark.

Furthermore, CUDA Thrust is a powerful library\cite{Thrust} which 
provides a rich collection of data parallel primitives such as
sort, scan, reduction, etc. Hence, using CUDA Thrust to build GPU applications,
programming efforts can be reduced greatly. 
Though CUDA Thrust does not support the {\tt sortperm}-like function directly,
we can simply use the {\tt sequence()} and {\tt sort\_by\_key()}
to implement it on GPU quickly, as shown in Figure~\ref{GPU_sortperm}.

\FloatBarrier

\begin{figure}[h]
\centerline{}
\scriptsize
\centering
\begin{Verbatim}[frame=single]
extern "C" void
sortperm_thrust(double *h_s, long long *h_idx, long long n)
{
   thrust::device_ptr<long long> d_idx = 
                     thrust::device_malloc<long long>(n);
   // create an array with elements 1, 2, 3, ..., n
   thrust::sequence(d_idx, d_idx + n, 1); 

   thrust::device_ptr<double> d_s = 
                     thrust::device_malloc<double>(n);

   thrust::copy(h_s, h_s+n, d_s); // copy s to device

   thrust::sort_by_key(d_s, d_s + n, d_idx);

   // we are interested in idx 
   thrust::copy(d_idx, d_idx + n, h_idx);

   thrust::device_free(d_s);
   thrust::device_free(d_idx);
}
\end{Verbatim}
\caption{Implementation of sortperm() on GPU}
\label{GPU_sortperm}
\end{figure}

\subsection{Sparse Matrix Multiplication and Addition}
\label{subsec:sparsmatmul}

For efficiently computing the matrix reduction, the Eirene library
uses the Schur complement to encapsulate the LU 
factorization \cite{HenselmanThesis}. 
Given a matrix M with submatrices A, B, C, and D as shown below,
\[
   M = \begin{bmatrix}
        A  & B \\
        C  & D 
       \end{bmatrix}
\]
the Schur complement S of the block A is
\[
   S = D + C A^{-1} B
\]
using the modulo-2 operation. Inside the Schur complement
function {\tt schurit4!()} in Eirene, the function {\tt blockprodsum()}
is invoked to compute $ D + C E $ after getting $ E = A^{-1} B $.  
Though Eirene uses sparse matrices for computing,
we found that the function {\tt blockprodsum()} becomes a bottleneck
for large arrays.

\begin{figure}[h]
\centerline{\includegraphics[width=3.1in]{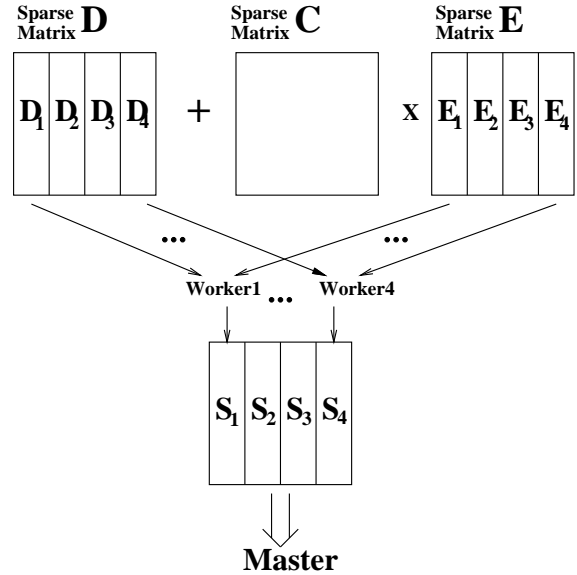}}
\caption{Parallel Implementation of blockprodsum() using Master/Workers}
\label{BlkProdSum}
\end{figure}

To alleviate this problem, we decided to take the advantage of the
multicore architecture and adopted the master/workers
parallel computation model to speed up the execution. 
We partitioned the matrices D and E
columnwise (due to the use of Compressed Sparse Column(CSC) format in Eirene)
into several workers (i.e. pthreads) and let each worker $i$ 
compute  $ S_i = D_i + C E_i $,
as illustrated in Figure~\ref{BlkProdSum}. Unlike the dense 
matrix multiplication in which the product matrix
is fully data parallel after partitioned, 
the index pointers in CSC which point to the starting locations of every column
have to be adjusted one after the other. We let the master do the adjusting
work when it copies the result to its caller.  

\section{Experimental Results}

To evaluate the effects of the performance-improving methods
discussed in the previous section, we ran  experiments
with three different versions of Eirene: the original version which
is Eirene v0.3.5 released in January 2017; the modified version
which removes the unnecessary array of type Any, avoids unnecessary calculations,
and uses RadixSort in {\tt sortperm()}; the enhanced version which
is a superset of the modified version and  
utilizes manycore/multicore to calculate the {\tt getstartweights\_subr2()},
{\tt sortperm()}, and the {\tt blockprodsum()} (using 4 workers). 

We firstly adopted the workstation at the Ohio Supercomputing Center(OSC)
to conduct the experiments. The machine has the Intel Xeon E5-2680 v4
CPU (2.4GHz), 28 cores per node, 128 GB of memory, as well as a 
cutting-edge NVIDIA Pascal P100 GPU(1.33GHz, 3584 CUDA cores, 16GB)
running CUDA Driver Version 8.0. Tables ~\ref{tab1},~\ref{tab2}, and ~\ref{tab3}
show execution times for the major bottlenecks and the total execution time
for the HIV, WorldMap and Dragon2 benchmarks, respectively. 
In these three tables, we use A, B, C, D, and E to denote the
major bottlenecks caused by abstract-type "Any" , 
{\tt integersinsameorderbycolumn()}, 
{\tt getstartweights\_subr2()}, {\tt sortperm()}, and {\tt blockprodsum()}.
   
\begin{table}[htbp]
\caption[caption]{Execution Times (in seconds) of the Major Bottlenecks using the HIV Benchmark\hspace{\textwidth} (Intel Xeon E5-2680 v4 and NVIDIA Tesla P100 (Pascal))}
\begin{center}
\begin{tabular}{|c||c||c||c|}
\hline
                             &  Original &  Modified  &  Enhanced  \\ \hline
A          &   95.7    &   0         &   0         \\ \hline
B          &   0.078   &  0.002      & 0.002        \\ \hline
C       &   3.2     &  3.2       &  0.068(Manycore)          \\ \hline
D                     &   0.42    &  0.043(RadixSort)   &  0.008(Manycore)          \\ \hline
E                &   0.135   &  0.135       &  0.134(Multicore)           \\ \hline
Total                        &   110.8   &  13.1        &  9.9          \\ \hline
\end{tabular}
\label{tab1}
\end{center}
\end{table}


\begin{table}[htbp]
\caption[caption]{Execution Times (in seconds) of the Major Bottlenecks using the WorldMap Benchmark\hspace{\textwidth} (Intel Xeon E5-2680 v4 and NVIDIA Tesla P100 (Pascal))}
\begin{center}
\begin{tabular}{|c||c||c||c|}
\hline
           &  Original &  Modified  &  Enhanced  \\ \hline
A          &  18.6    &     0      &    0       \\ \hline
B          &   2.2   &     0.002    &   0.002      \\ \hline
C          &   1.3    &    1.3     &    0.29(Manycore)        \\ \hline
D          &   38.8    &   4.1 (RadixSort)   &  0.39(Manycore)           \\ \hline
E          &   1.18   &    1.18     &    0.90(Multicore)         \\ \hline
Total      &   80.6   &  23.4       & 17.1    \\ \hline
\end{tabular}
\label{tab2}
\end{center}
\end{table}


\begin{table}[htbp]
\caption[caption]{Execution Times (in seconds) of the Major Bottlenecks using the Dragon2 Benchmark\hspace{\textwidth} (Intel Xeon E5-2680 v4 and NVIDIA Tesla P100 (Pascal))}
\begin{center}
\begin{tabular}{|c||c||c||c|}
\hline
           &  Original &  Modified  &  Enhanced  \\ \hline
A          &   416.6   &    0        &   0         \\ \hline
B          &    36.2   &    0.05    &    0.05     \\ \hline
C          &    15.4   &    15.4     &  0.72(Manycore)          \\ \hline
D          &    2.4   &  0.41(RadixSort)    &  0.03(Manycore)           \\ \hline
E          &    15.5  &  15.5       &   9.6(Multicore)          \\ \hline
Total      &   565.2   &  109.9        &  89.0 \\ \hline
\end{tabular}
\label{tab3}
\end{center}
\end{table}

It can be seen that the removal of unnecessary use of the array of type {\tt Any}
can greatly improve the performance for all benchmarks. 
The other methods also have positive
impacts on the performance for different benchmarks. 
Note that there is no significant improvement
for the HIV and Worldmap benchmarks when using multicore to speed up
the {\tt blockprodsum()} due to the small amount of computation.
However, we used different number of threads to compute the 
{\tt blockprodsum()} for the Dragon2
benchmark.  As shown in Table~\ref{workers}, using a few more threads
can still  improve the performance.      
For parallel dense matrix multiplication, usually linear or close to 
linear speedups can be obtained.
The reason we cannot get
linear speedups here is because the matrices are sparse and the workload
is not fully balanced among the worker threads.  
Currently, a load balancing implementation of the {\tt blockprodsum()}
function is being developed.

\begin{table}[htbp]
\centerline{}
\centerline{}
\caption[caption]{Execution Times (in seconds) of the blockprodsum() in Dragon2 Benchmark with different number of workers }
\begin{center}
\begin{tabular}{|c|c|c|c|c|c|c|c|}
\hline
Num. of Workers  &  1    & 2    &  4   &  6  &  8 & 10 & 12 \\  \hline \hline
Time             &  15.5 & 11.3 & 9.6  & 9.2 & 8.6  & 8.4 & 8.3  \\ \hline
\end{tabular}
\label{workers}

\end{center}
\end{table}

In addition to the three benchmarks mentioned above, we also ran another four
benchmarks: Dragon1, C.\ elegans, Klein\_4, and Klein\_9, all available
from \cite{OtterRoadMap}. The Dragon1 is a 1000-point data set sampled
from the Stanford Dragon graphic, the C.\ elegans benchmark is a neuronal
network with 297 neurons, while Klein\ 400 and Klein\ 900 are the data sets
sampled from the Klein bottle pictures with 400 and 900 points, respectively.  
Except the WorldMap benchmark which sets epsilon = 0.15, all benchmarks use Eirene's
default setting Inf as the max distance threshold.
Table~\ref{OSCmach} shows the timing results on a workstation at OSC.
To show our methods can also work well on different hardware platforms,
we ran the benchmarks using the workstation in our laboratory. 
Table~\ref{CSUmach} displays
the results from a workstation with Intel Xeon CPU E3-1231(3.40GHz, 8GB memory) 
and NVIDIA Quadro K620 (Maxwell) GPU. This machine has higher
CPU clock rate but older GPU model than the OSC workstation. 

\begin{table}[htbp]
\caption[caption]{Execution Times (in seconds) of the Benchmarks using Intel Xeon E5-2680 v4 and NVIDIA Tesla P100 (Pascal)}
\begin{center}
\begin{tabular}{|c|r|r|r|}
\hline
Name (Size of complex)    &  Original    & Modified  &  Enhanced \\ \hline \hline
C.\ elegans ($4.4\times10^6$)  &  4.9        &  2.2     &   1.9          \\ \hline \hline
Klein\ 400 ($1.1\times10^7$)   &  7.4      &    2.7    &    2.3     \\ \hline \hline
Klein\ 900 ($1.2\times10^8$)   &  61.9      &    10.3    &    9.8     \\ \hline \hline
HIV        ($2.1\times10^8$)   &  110.8      &   13.1    &    9.9     \\ \hline \hline
Dragon1    ($1.7\times10^8$)   &  57.3      &    12.8    &   11.2     \\ \hline \hline
Dragon2    ($1.3\times10^9$)   &  565.2      &   109.9    &  89.0     \\ \hline \hline
WorldMap   ($2.0\times10^{8}$)  &  80.6      &     23.4    &  17.1     \\ \hline
\end{tabular}
\label{OSCmach}
\end{center}
\centerline{}
\end{table}

\begin{table}[htbp]
\caption[caption]{Execution Times (in seconds) of the Benchmarks using Intel Xeon E3-1231 and NVIDIA Quadro K620 (Maxwell)}
\begin{center}
\begin{tabular}{|c|r|r|r|}
\hline
Name (Size of complex)    &  Original    & Modified  &  Enhanced \\ \hline \hline
C.\ elegans ($4.4\times10^6$)  &  4.7        &  2.2     &   2.1          \\ \hline \hline
Klein\ 400 ($1.1\times10^7$)   &  7.3      &    2.6    &    2.5     \\ \hline \hline
Klein\ 900 ($1.2\times10^8$)   &  58.2      &    11.5    &    10.1     \\ \hline \hline
HIV        ($2.1\times10^8$)   &  108.4      &   13.0    &    11.1     \\ \hline \hline
Dragon1    ($1.7\times10^8$)   &  54.1      &    12.4    &   11.7     \\ \hline \hline
Dragon2    ($1.3\times10^9$)   &  543.0      &   110.3    &  94.0     \\ \hline \hline
WorldMap   ($2.0\times10^{8}$)  &  74.9      &     23.1    &  17.9     \\ \hline
\end{tabular}
\label{CSUmach}
\end{center}
\centerline{}
\end{table}

We also measured the memory usage(in GB) of each benchmark on 
two different hardware platforms. 
We found that the memory usage stays constant on different machines,
due to the same software (i.e. Julia) version and configuration.
Therefore, we only present one set of data in Table\ref{MemoryUsage}
which shows the size of memory used by each benchmark under different 
implementations of Eirene.
It can be observed that the original version of Eirene used much more memory
because it used the array of the abstract-type Any. The enhanced version
used a little more memory than the modified version due to some memory
allocated in the wrapper functions. 

\begin{table}[htbp]
\caption[caption]{Memory Usage (in GB) for each version of the Benchmarks on three Different Hardware Platforms $\;\;\;\;\;\;\;\;\;\;\;\;$  }
\begin{center}
\begin{tabular}{|c|r|r|r|}
\hline
Name (Size of complex)    &  Original    & Modified  &  Enhanced \\ \hline \hline
C.\ elegans ($4.4\times10^6$)  &  1.30        &  0.61     &  3.96        \\ \hline \hline
Klein\ 400 ($1.1\times10^7$)   &  2.43      &    1.07    &   4.43     \\ \hline \hline
Klein\ 900 ($1.2\times10^8$)   &  28.91      &    7.23    &   13.80     \\ \hline \hline
HIV        ($2.1\times10^8$)   &  43.55      &   9.75   &    13.02     \\ \hline \hline
Dragon1    ($1.7\times10^8$)   &  24.80      &    6.94   &   11.30     \\ \hline \hline
Dragon2    ($1.3\times10^9$)   &  262.59      &   51.02    &  56.91    \\ \hline \hline
WorldMap   ($2.0\times10^{8}$)  & 22.99       &  15.88    &  21.96    \\ \hline
\end{tabular}
\label{MemoryUsage}
\end{center}
\centerline{}
\end{table}

\section{Conclusion}

We used the profiling tools in Julia to identify the bottlenecks in
Eirene, an open-source platform for computing persistent homology.
Several performance-improving methods targeting the bottlenecks
have been developed, such as removing unnecessary use of the array of the
abstract-type Any,
eliminating redundant computation, use of manycore/multicore to accelerate
execution, etc. Experimental results demonstrate that the performance
can be greatly improved. 

\centerline{}
\centerline{}

\section*{Acknowledgment}
This research was supported by the NASA GRC Summer Faculty Fellowship
and by allocation of computing time from the Ohio Supercomputer Center. 

\pagebreak

\end{document}